\documentclass{article}

\usepackage{arxiv}

\usepackage[utf8]{inputenc} 
\usepackage[T1]{fontenc}    
\usepackage{hyperref}       
\usepackage{url}            
\usepackage{booktabs}       
\usepackage{amsfonts}       
\usepackage{nicefrac}       
\usepackage{microtype}      
\usepackage{lipsum}
\usepackage{color}
\usepackage{algorithm}
\usepackage{algorithmic}
\usepackage{amsmath,amsfonts,amssymb}
\usepackage{subfigure}
\usepackage{lineno}
\usepackage{graphicx}
\usepackage{bm}
\usepackage{subfigure}

\title{Data-driven nonlinear turbulent flow scaling with Buckingham Pi variables}

\author{
Kai Fukami$^{[1,*]}$, Susumu Goto$^{[2]}$, Kunihiko Taira$^{[1]}$
\\
\\
1. Department of Mechanical and Aerospace Engineering,
University of California, Los Angeles, CA 90095, USA\\
2. Graduate School of Engineering Science, Osaka University, 1-3 Machikaneyama, Toyonaka, Osaka, 560-8531, Japan\\
Corresponding author: kfukami1@g.ucla.edu
}

\begin{document}
\maketitle

\begin{abstract}

Nonlinear machine learning for turbulent flows can exhibit robust performance even outside the range of training data.  
This is achieved when machine-learning models can accommodate scale-invariant characteristics of turbulent flow structures. 
This study presents a data-driven approach to reveal scale-invariant vortical structures across Reynolds numbers that provide insights for supporting nonlinear machine-learning-based studies of turbulent flows.  
To uncover conditions for which nonlinear models are likely to perform well, we use a Buckingham Pi-based sparse nonlinear scaling to find the influence of the Pi groups on the turbulent flow data.  
We consider nonlinear scalings of the invariants of the velocity gradient tensor for an example of three-dimensional decaying isotropic turbulence.  
The present scaling not only enables the identification of vortical structures that are interpolatory and extrapolatory for the given flow field data but also captures non-equilibrium effects of the energy cascade.
As a demonstration, the present findings are applied to machine-learning-based super-resolution analysis of three-dimensional isotropic turbulence.  
We show that machine-learning models reconstruct vortical structures well in the interpolatory space with reduced performance in the extrapolatory space revealed by the nonlinearly scaled invariants.  
The present approach enables us to depart from labeling turbulent flow data with a single parameter of Reynolds number and comprehensively examine the flow field to support training and testing of nonlinear machine-learning techniques.

\end{abstract}

\section{Introduction}
\label{sec:intro}

Trained fluid mechanicians can identify similarities in vortical structures for a variety of turbulent flows.  
Even if there are scale or rotational differences, we can visually extract similar structures due to their geometrical features across spatial and temporal scales.
Analogously, recent machine-learning models have capitalized on such structural similarities to achieve reliable performance for the given training data.  
However, it is also known that machine-learning models often achieve satisfactory performance even in the cases of unseen Reynolds numbers.  
Examples of models performing successfully beyond the training data include, turbulence modeling \cite{guan2021stable}, state estimation \cite{guastoni2021convolutional}, and super-resolution \cite{kim2021unsupervised}.
This suggests that machine-learning models are able to incorporate scale-invariant properties of the flows while optimizing the output to meet their objectives.

\begin{figure}
    \centering
    \includegraphics[width=0.87\textwidth]{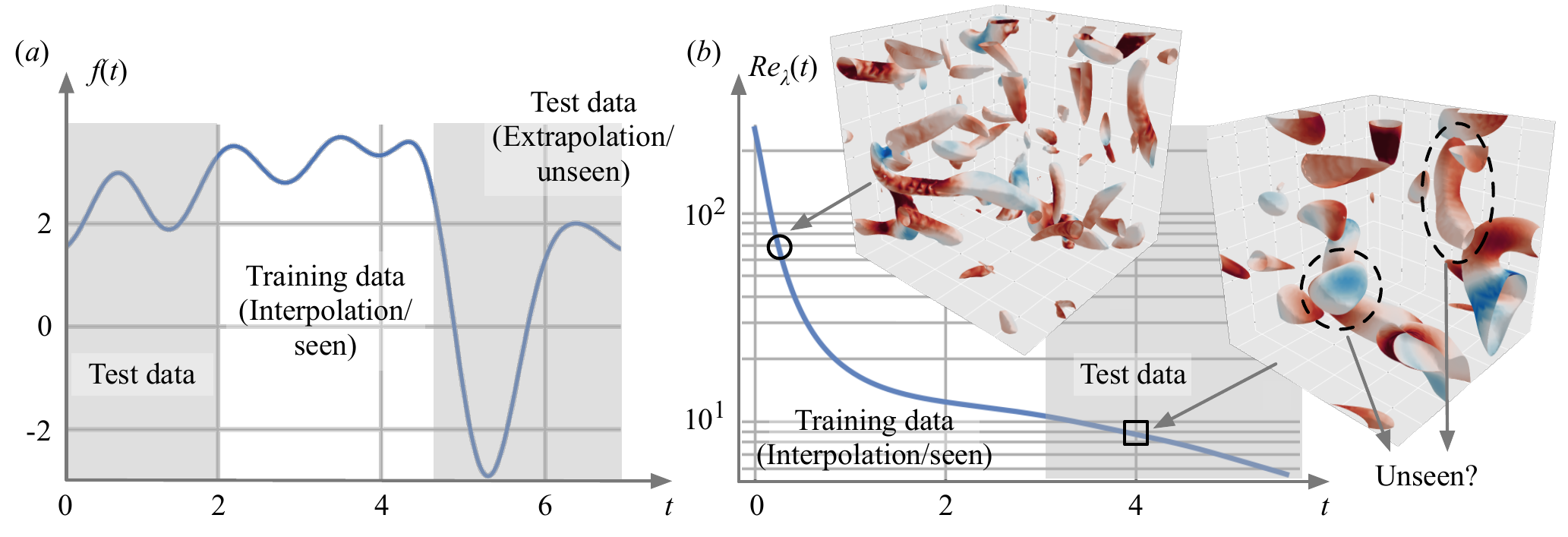}
    
    \vspace{-3mm}
    \caption{
    Concept of interpolation and extrapolation.
    $(a)$ A one-dimensional example.
    $(b)$~Evolution of three-dimensional decaying isotropic turbulence over $Re_\lambda(t)$.
    }
    \vspace{-3mm}
    \label{fig1}
\end{figure}

These observations imply that nonlinear machine-learning models are capturing data characteristics in a more holistic manner.  
Traditionally, interpolation and extrapolation of models have been associated with a particular variable or parameter~\cite{domingos2012few,marcus2018deep}, as illustrated in figure~\ref{fig1}$(a)$.  
In this example, test data outside of the training data is captured by the temporal variable $t$.  
In the case of more complex data set, such as turbulent flows, describing whether a certain type of vortical structure appears in training data or not requires additional considerations, beyond a single parameter such as the Reynolds number.

Given this motivation, we revisit the idea of interpolation and extrapolation in the context of turbulent flow structures.  
Here, we consider the capturing vortical structures that have certain similarities across training and test data to be ``interpolation'' and their structural features to be ``seen.''  
On the other hand, capturing structures in the test data that do not share similarities with those in the training data are referred to as ``extrapolation'' and their structures as ``unseen.''
With machine-learning models known to perform well for approximating the nonlinear relationship between the input and output from a large collection of data, the presence of seen (or common) features can provide robust performance under untrained flow situations \cite{kim2021unsupervised,guan2021stable}. 
The current study presents a data-driven scaling approach to reveal seen/unseen structures of turbulent flow data in machine learning.

To assess rotational and shear similarities between small- and large-length scales in turbulent flows, we examine the flow field data in terms of the invariants of the velocity gradient tensor.
In the present analysis, the scaled invariants are found through sparse nonlinear regression using non-dimensional parameters from the Buckingham Pi theorem \cite{buckingham1914physically}.
The present approach offers the optimal nonlinear scalings of the invariants, uncovering scale-invariant vortical structures in a turbulent flow field.
Analyzing the data distribution of the present scaled invariants reveals what types of flow structures are seen and unseen.  
Furthermore, the findings from this study provide guidance in the choice of machine-learning functions to offer robustness for scale-invariant vortical structures. 
The present paper is organized as follows.  
The proposed Buckingham-Pi sparse nonlinear scaling of the invariants is introduced in section~\ref{sec:method}.  
We demonstrate the current approach for three-dimensional isotropic turbulence in section~\ref{sec:apriori}.  
Concluding remarks are provided in section~\ref{sec:conc}.

\section{Methods}
\label{sec:method}

\vspace{-0.5mm}
Machine-learning models for turbulent flow structures are known to remain accurate beyond the coverage of training data.  
This is the case especially when the nonlinear machine-learning models have scale and rotational invariances embedded in their formulations.  
To explain this extended validity of machine-learning models, this study aims to uncover nonlinear scalings that capture the similarities in turbulent vortical structures across a range of Reynolds numbers.  
The central hypothesis of this study is that the existence of such scalings enables nonlinear machine-learning techniques to effectively perform across different flow fields beyond the range of Reynolds numbers provided in the training data.  
With these identified scalings, turbulent flow structures that linearly and nonlinearly span across a range of scales may be considered {\it seen} beyond the training data due to their structural similarities.

To examine turbulent flows, we consider the invariants of the velocity gradient tensor $\bm{A} = (\nabla {\bm u})^T$ such that the observations are independent of the frame of reference~\cite{chong1990general}.  
These invariants are $P = \text{trace}(\bm{A})$, $Q = \frac{1}{2} [ P^2 - \text{trace}(\bm{A}^2) ]$, and $R = \text{det}(\bm{A})$.  
For the present study, we consider incompressible turbulent flows, which makes $P = \nabla\cdot{\bm u} = 0$.  
The remaining two invariants of $Q$ and $R$ characterize the local rotation and shear, respectively~\cite{ooi1999study}.
According to these invariants, the flow can experience vortex compression $(Q>0, R<0)$, vortex stretching $(Q>0, R>0)$, biaxial strain $(Q<0, R<0)$, and axial strain $(Q<0, R>0)$~\cite{davidson2015turbulence}.
These invariants will be nonlinearly scaled with non-dimensional Pi groups using a data-driven approach.

\begin{figure}
    \centering
    \includegraphics[width=\textwidth]{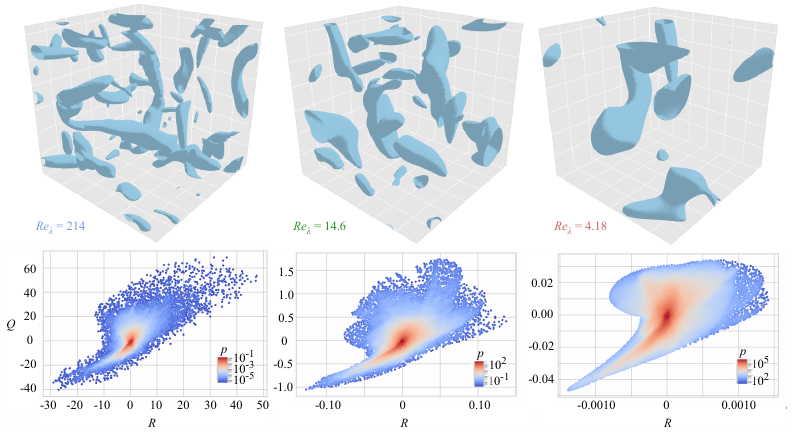}
    
    \vspace{-2mm}
    \caption{
    Example flow snapshots with $Q-R$ distributions of three-dimensional decaying turbulence at $(a)$ $Re_{\lambda} = 214$,  $(b)$~$14.6$, and $(c)$~$4.18$.
    Each distribution is colored by density.
    Turbulent vortices are visualized with $(a)$ $Q = 10$, $(b)$~$0.3$, and $(c)$~$0.02$.
    }
    \vspace{-4mm}    
    \label{fig2}
\end{figure}

In this study, we consider three-dimensional incompressible decaying isotropic turbulence.  
The flow field data is obtained from direct numerical simulation using $64^3$ grid points with a Taylor microscale-based Reynolds number of {$0.85 \leq Re_{\lambda} \leq 252$}~\cite{chumakov2008priori,meena2021identifying}, satisfying $k_{\rm max}\eta \geq 1$, where $k_{\rm max}$ is the maximum resolvable wavenumber of the grid and $\eta$ is the Kolmogorov length scale, to resolve all important scales of motion.
The present simulation is initialized with random velocity fields of Gaussian profiles that satisfy incompressibility and the Kolmogorov spectra for kinetic energy.
The flow snapshots are curated after the flow reaches the decaying regime at which we set time $t$ to zero.
Note that a large $Re_\lambda$ at the early stage of decay is due to a small dissipation coefficient, which strongly depends on the flow~\cite{goto2009dissipation,goto2015energy}.
We present representative flows with their corresponding invariants on the $Q$--$R$ plane over time in figure~\ref{fig2}.  
The Reynolds number decreases as vortical structures evolve as their characteristic size increases.  
Even at a large time when $Re_\lambda = 4.18$, large-scale structures are still observed while the flow is under decay.
During this process, invariants $Q$ and $R$ decrease in magnitude while the probability density functions of these invariants remain geometrically similar.  
This suggests that there is some level of scale invariance in the distributions of the turbulent vortical structures over the $Q$--$R$ plane.  
That is, the decaying isotropic turbulence holds similar rotational and shear structures whose sizes vary over the Reynolds number.

\begin{figure}
    \centering
    \includegraphics[width=0.45\textwidth]{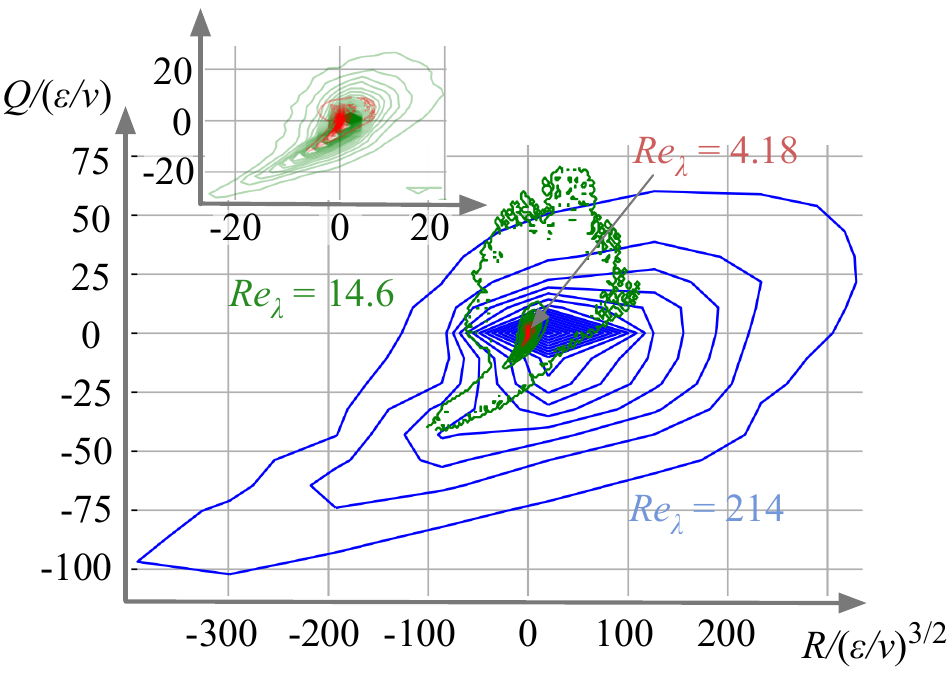}
    \caption{
    Linearly-scaled $Q$--$R$ distributions based on the kinematic viscosity $\nu$ and energy dissipation rate $\epsilon$.
    A zoom-in view for $Re_\lambda = 14.6$ and 4.18 is also shown. 
    }
    \label{fig3-new}
\end{figure}

Linear scaling for the $Q$--$R$ distributions based on the kinematic viscosity $\nu$ and energy dissipation rate $\epsilon$ such as $Q/(\epsilon/\nu)$ and $R/(\epsilon/\nu)^{3/2}$ do not completely collapse the distributions, given their long tails especially at high Reynolds number, as presented in figure~\ref{fig3-new}.
This seems to be caused by the wide range of Reynolds numbers and strong unsteadiness contained in the present decaying turbulence data, implying that Kolmogorov's similarity hypotheses do not hold in an instantaneous manner.
For example, skewness $S_R$ of the linearly-scaled distribution for $R$ in figure~\ref{fig2} is 2.74, 0.299, and 0.186 for $Re_{\lambda}=214$, 14.6 and 4.18, respectively.
These observations suggest that nonlinear scaling needs to be considered to accommodate non-equilibrium effects.
Identifying such nonlinear scalings of $Q$ and $R$ distributions with data-driven techniques can also reveal how nonlinear machine learning extracts {\it seen} (or {\it common}) vortical structures that share similarities with structures outside of the training data sets.  
Moreover, {\it unseen} structures can be captured by uncovering the invariant space that do not overlap for the scaled data.

Let us consider scaling the invariants $Q$ and $R$ in a nonlinear manner using non-dimensional Pi groups from the Buckingham Pi theorem~(1914), which distills a number of dimensional parameters into a smaller number of dimensionless groups~\cite{xie2022data,Joe2022BP}.
We assume that the scaling can be obtained through superposing appropriate polynomials of the Pi variables.  
Denoting these invariants as $\phi$ (either $Q$ and $R$), the candidate polynomials are assumed to have the form of
\vspace{-1mm}
\begin{equation}
    {\theta}_k(\bm{x},t) = \Pi_i^m(t) \Pi_j^n(t) \phi(\bm{x},t), 
    ~~ 
    \text{where}
    ~~
    i,j = 1, 2, \dots,
    ~~
    m, n \in \mathbb{Z},
    ~~
    {k = 1, \dots, n_L}.
    \label{eq:cand_poly}
\end{equation}
Here, $n_L$ is the number of library-basis functions.
Given these library candidates $\theta_k$, we can express the scaled invariants ${\phi}^*(\bm{x},t)$ as
\vspace{-3mm}
\begin{align}
    {\phi}^*(\bm{x},t) = \sum_{k=1}^{n_L} a_k \theta_k(\bm{x},t)
\end{align}
with ${\bm a} = (a_1,a_2, \dots, a_{n_L})^T \in \mathbb{R}^{n_L}$.
Spatiotemporal discretization of this equation yields ${\bm \Phi}^* = \bm{\Theta} \bm{a},$ where
\vspace{-2mm}
\begin{multline}
    {\bm \Phi}^* =
    \begin{bmatrix} 
    \bm{\phi}^*(t_1) \\
    \bm{\phi}^*(t_2) \\
    \vdots \\
    \bm{\phi}^*(t_{n_t}) 
    \end{bmatrix}
    \in \mathbb{R}^{n_x n_t}
    ,\quad
    {\bm \Theta} = 
    \begin{bmatrix} 
    \bm{\theta}_1(t_1) & \bm{\theta}_2(t_1) & \dots & \bm{\theta}_{n_L}(t_1) \\
    \bm{\theta}_1(t_2) & \bm{\theta}_2(t_2) & \dots & \bm{\theta}_{n_L}(t_2) \\
    \vdots & \vdots & \ddots & \vdots \\
    \bm{\theta}_1(t_{n_t}) & \bm{\theta}_2(t_{n_t}) & \dots & \bm{\theta}_{n_L}(t_{n_t}) \\
    \end{bmatrix} \in \mathbb{R}^{n_x n_t \times n_L}
\end{multline}
with ${\bm\phi}(t) = (\phi(x_1,t), \dots, \phi(x_{n_x},t))^T \in \mathbb{R}^{n_x}$ and ${\bm\theta}_k(t) = (\theta_k(x_1,t), \dots, \theta_k(x_{n_x},t))^T \in \mathbb{R}^{n_x}$ being the scaled invariant and the library candidate, respectively, discretized in space. 
Here, the invariants are spatiotemporally stacked into a tall vector.  
The coefficients ${\bm a}$ are determined through the sequential threshold least squares method~\cite{brunton2016discovering}, promoting sparsity of the coefficient matrix in a computationally efficient manner~\cite{fukami2020sparse}.
Since excessive sparsity promotion leads to a high regression error, the sparsity coefficient in the threshold least square method needs to be carefully tuned~\cite{kaiser2018sparse}.

For the present decaying turbulence, the Taylor length scale $\lambda$ can be expressed as the function of the characteristic velocity $u$ (the square root of the spatially averaged kinetic energy), the kinematic viscosity $\nu$, the computational domain size $L$, and viscous dissipation $\epsilon$ such that $\lambda = f(u,\nu, L, \epsilon)$.
Through the use of Buckingham Pi theorem, we can find three Pi variables, which namely are $\Pi_1 = u\lambda/\nu$, $\Pi_2 = uL/\nu$, and $\Pi_3 = \epsilon \nu /u^4$.  
The first Pi variable $\Pi_1$ is the Taylor Reynolds number $Re_\lambda$.
In the present study, we take these Pi variables as a function of time.  
Given these Pi variables, we can construct the library candidates for equation~(\ref{eq:cand_poly}) as first and second-order polynomials of the Pi variables such that
\vspace{-1mm}
\begin{equation}
\begin{split}
    \{ \Pi_1,~\Pi_2,&~\Pi_3,~ 
    \Pi_1^{-1},~\Pi_2^{-1},~\Pi_3^{-1},~
    \Pi_1^2,~\Pi_2^2,~\Pi_3^2,~
    \Pi_1^{-2},~\Pi_2^{-2},~\Pi_3^{-2},\\
    &\Pi_1\Pi_2,~ \Pi_2 \Pi_3,~ \Pi_3 \Pi_1,~ 
    \Pi_1^{-1}\Pi_2,~ \Pi_2^{-1} \Pi_3,~ \Pi_3^{-1} \Pi_1,
    \dots,~ \Pi_i^m \Pi_j^n,~ \dots \},
\end{split}
\end{equation}
where $|m|+|n|\le 2$.

With the Buckingham Pi-based library matrix ${\bm \Theta}$, we seek coefficients $\bm a$ by maximizing the similarity of data distributions of the invariants over space and time.
Here, we utilize the Kullback–Leibler (KL) divergence~\cite{kullback1951information} to assess the difference between a probability distribution $f_1$ and the reference probability distribution $f_2$.
For two probability (data) distributions, the KL divergence is defined as 
\vspace{-1mm}
\begin{align}
    D(f_{1}||f_{2}) 
    \equiv \int_{-\infty}^{\infty} f_{1}({\bm\phi}^*) 
    {\rm log}\dfrac{f_{2}({\bm\phi}^*)}{f_{1}({\bm\phi}^*)} {\rm d}{\bm\phi}^*.
    \label{eq:D}
\end{align}
The minimization of the KL divergence finds the optimal coefficients ${\bm a}^*$ for maximum similarity of the scaled invariant distributions, which yields an optimization problem of 
${\bm a}^* = {\rm argmin}_{\bm a}[D(f_1||f_2)].$
Below, the data distribution from a snapshot at high $Re_\lambda$
is used as the reference 
$f_2$.

\section{Results}
\label{sec:apriori}

\vspace{-0.5mm}
We apply the present nonlinear scaling analysis to three-dimensional decaying isotropic turbulence over $0.85 \leq Re_{\lambda} \leq 252$.
The present data is comprised of 800 snapshots over time and $64^3$ grid points in a tri-periodic cubic domain.
Here, we aim to (1) uncover the nonlinear influence of characteristic variables on the evolution of the invariants $Q$ and $R$, (2) identify common (seen) vortical flow features through the scaled invariants, and (3) provide guidance on proper training and formulation of machine-learning models.

\begin{figure}
    \centering
    \includegraphics[width=\textwidth]{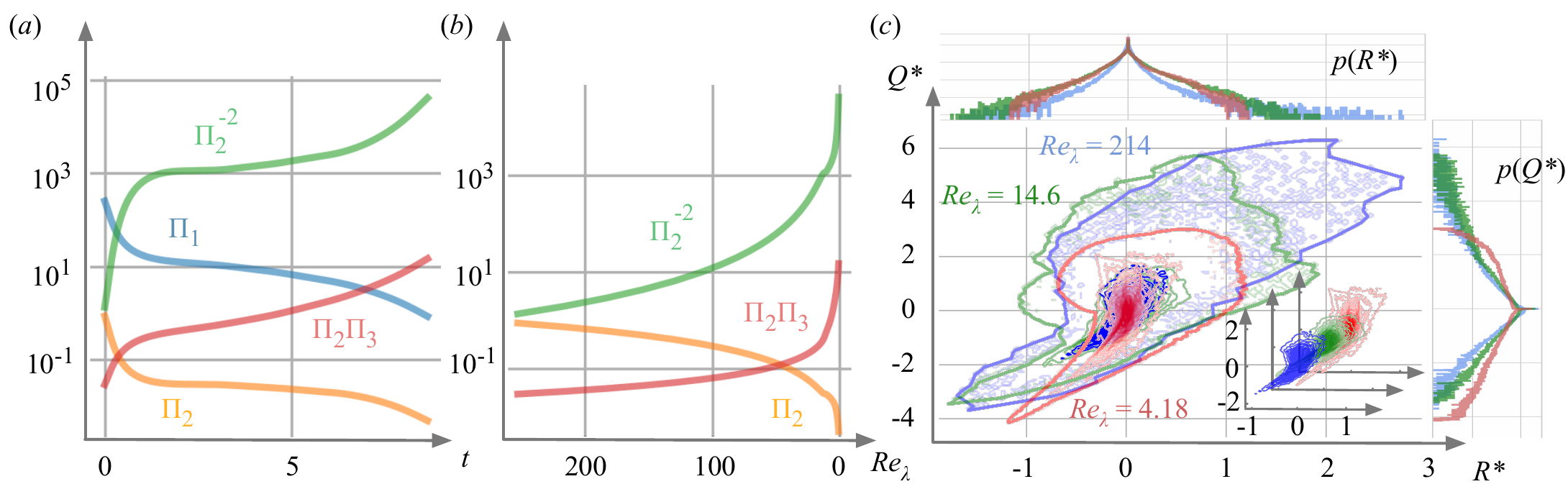}
    \vspace{-4mm}
    \caption{
    The data-driven Buckingham Pi scaling for three-dimensional decaying turbulence.    
    The time series of the identified non-dimensional variables as a function of $(a)$ time and $(b)$ the Taylor Reynolds number.
    $(c)$~Scaled $Q^*$ and $R^*$ invariants with their probability density functions.
    A zoom-in view is also shown as a subfigure.
    }
    \vspace{-3mm}
    \label{fig3}
\end{figure}

Based on sparse regression, we find the scaling factors for $Q$ and $R$ to be
\begin{align}
    \label{3Dq}
    &Q^* = (5.76\Pi_1+4.17\Pi_2^{-2}-3.59\Pi_2\Pi_3)Q,\\\label{3Dr}
    &R^* = (5.53\Pi_1-0.826\Pi_2^{-1}+4.42\Pi_2^{-2}-3.44\Pi_2\Pi_3)R.
\end{align}
While isotropic turbulence is complex, the nonlinearly scaled invariants turn out to be surprisingly compact in their expressions.
For both scaled invariants, we have $\Pi_1$ which is the Taylor Reynolds number $Re_\lambda$.
This reflects the decaying nature, corresponding to the fact that the data spread on the $Q$--$R$ plane shrinks with decreasing Taylor Reynolds number over time, as shown in figures~\ref{fig3}$(a)$ and $(b)$.
The present scaling also identifies the influence of $\Pi_2= uL/\nu$, the box-size-based Reynolds number, 
revealing that the computational domain size influences the turbulent flow, especially as time advances and vortical structures become comparable in size to the computational domain.
This is evident from the inverse and quadratic inverse nature of the scalings.
The size of the periodic box affects not only the large-scale vortical motion over the domain but also the energy dissipation of turbulence~\cite{davidson2015turbulence}.

Furthermore, the present analysis uncovers the importance of $\Pi_2\Pi_3 = uL/\nu \cdot \epsilon \nu /u^4 = \epsilon L/u^3$, which is the ratio between the instantaneous energy dissipation rate determined by small length scales and the cascading energy flux from the system-size vortices.
Thus, $\Pi_2\Pi_3$ plays a similar role to the dissipation coefficient $C_\epsilon \equiv \epsilon l/u^3$, quantifying the degree of non-equilibrium~\cite{vassilicos2015dissipation}, i.e., the violation of Kolmogorov's similarity.
Although $\Pi_2\Pi_3$ is constant when the energy flux of large-scale vortices and energy dissipation rate are in balance, that is not the case for the present decaying turbulence~\cite{goto2016local}.
The present $\Pi_2\Pi_3$ accounts for the time delay of energy-cascade process between energy flux and dissipation due to strong nonlinearities in decaying turbulence.
This is achieved by correcting non-equilibrium effects in a nonlinear manner.
Note that the present formulation provides similar scaling expression in a range of similar Reynolds numbers even with different initial flow fields since sparse regression captures how the turbulent flow decays over time.

Next, the identified scalings are applied to the original $Q$ and $R$ data distributions, as shown in figure~\ref{fig3}$(c)$ along with their probability density functions.
The identified factors in equations~\ref{3Dq} and~\ref{3Dr} yield the optimal overlap of $Q^*$ and $R^*$ over space and time.
The scaled data distribution for the high Reynolds number flow spreads over a larger area than that for the low Reynolds number, especially for $Q^*>0$.
This implies high occurrence of vortex stretching and compression at high Reynolds number.

\begin{figure}
    \centering
    \includegraphics[width=0.85\textwidth]{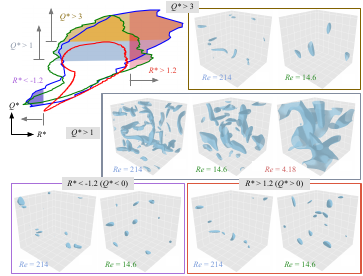}
    \caption{
    Interpolatory and extrapolatory vortical structures in three-dimensional decaying isotropic turbulence.  
    }
    \vspace{-5mm}
    \label{fig4}
\end{figure}

Let us focus on the vortical structures for overlapping and non-overlapping regions of the data over the $Q^*$--$R^*$ plane, as illustrated in figure~\ref{fig4}.
For $Q^*>1$, similar shapes of vortical elements can be observed across a range of length scales.
These identified vortical structures are assessed as `interpolatory' (common) with the present scaling approach.
The isosurfaces of $Q^*>3$ are also visualized in figure~\ref{fig4}.
These strong rotational elements that barely appear in low Reynolds number flows are `extrapolatory' structures.
Moreover, such `extrapolatory' vortical structures against the low-$Re$ flow field can be seen in the portion of $R^*<-1.2$ and $R^*>1.2$.
The region of $Q<0$ and $R<0$ corresponds to biaxial strain, while that of $Q>0$ and $R>0$ reflects vortex stretching~\cite{davidson2015turbulence}.
Hence, the scaled data distribution suggests that these structures caused by strong biaxial strain and vortex stretching correspond to `extrapolation' for the present turbulence data.
While not shown here, the portion of $R^*> -0.5$ and $-3.5<Q^*<0.5$ also includes extraporatory vortical structures.
The scaled $Q$--$R$ data distribution gives insights into turbulent flows, in addition to the identified scaling.

The present Buckingham-Pi-based sparse nonlinear scalings can identify interpolatory and extrapolatory vortical structures in isotropic turbulence.
Nonlinear machine-learning models are likely to capture the characteristics and behavior of what we refer to as the interpolatory structures even in untrained Reynolds number cases.  
This explains why well-trained machine-learning models may perform well even for testing data.  
However, we caution that when the test data includes extrapolatory structures, machine-learning models are no longer guaranteed to be valid.
With regard to these points, classifying interpolatory and extrapolatory structures solely by $Re_\lambda$ is not encouraged for assessing nonlinear machine-learning models.

Based on the insights from the scaled invariants above, let us consider machine-learning-based super-resolution reconstruction of turbulent flows~\cite{FFT2023_survey}.
Super resolution reconstructs the high-resolution flow field ${\bm q}_{\rm HR}$ from its low-resolution data ${\bm q}_{\rm LR}$ with a reconstruction model ${\cal F}$ through ${\bm q}_{\rm HR}={\cal F}({\bm q}_{\rm LR})$.
Recently developed machine-learning-based super-resolution analysis captures the nonlinear relationship between small (unresolved) and large-scale (resolved) structures.  
This study considers the ability of machine-learning-based reconstruction to recognize and reconstruct {\it common} turbulent flow structures for a variety of flow field snapshots, even outside of the training data.

For super-resolution reconstruction of turbulent flows, a machine-learning model should be carefully constructed to account for a range of spatial length scales while enforcing rotational and translational invariance of vortical structures.
To satisfy these properties, we use the hybrid downsampled skip-connection/multi-scale (DSC/MS) model~\cite{FFT2019a,FFT2021}.
The DSC/MS model based on convolutional neural network~\cite{LBBH1998} is composed of three main functions: namely, (A) up/downsampling operations, (B) skip connection, and (C) multi-scale filters.
The up/downsampling provides robustness against rotation and translation of vortical structures.
The skip connection allows communication between the input low-resolution data and the intermediate output of the DSC model, which is crucial in learning a step-by-step internal process towards the high-resolution output from the low-resolution input while expanding the dimension of the flow field snapshot~\cite{he2016deep}.
The multi-scale filters apply filtering with a number of different sizes of them (e.g., three here) in parallel to capture a broad range of scales in turbulent flows.
Further details on the present neural network model are in Fukami et al.~\cite{FFT2019a}.
We will discuss which function inside the present model contributes to gaining robustness for scale invariance later.

Here, we consider two cases of training and testing for super-resolution reconstruction of turbulent flows: (1) a model trained with a low-$Re_{\lambda}$ data set and tested with a high-$Re_{\lambda}$ data set (low-$Re_{\lambda}$ training); (2) a model trained with a high-$Re_{\lambda}$ data set and tested with a low-$Re_{\lambda}$ data set (high-$Re_{\lambda}$ training).
We expect that the low-$Re_{\lambda}$ training cannot cover the non-overlapping portion of the scaled invariants, which corresponds to extrapolation.
The high-$Re_{\lambda}$ training, which covers a wide portion of the scaled invariants, may amount to interpolation in terms of the turbulent flow structures.
For the present analysis with the two training scenarios, the threshold $Re_{\lambda}$ between the low- and high-$Re_{\lambda}$ training cases is set to $Re_{\lambda}=15$.
The velocity vector is used as data attributes $\bm q$ with $\cal F$ reconstructing the high-resolution flow field on $64^3$ grids from the low-resolution data on $4^3$ grid.

\begin{figure}
    \centering
    \includegraphics[width=\textwidth]{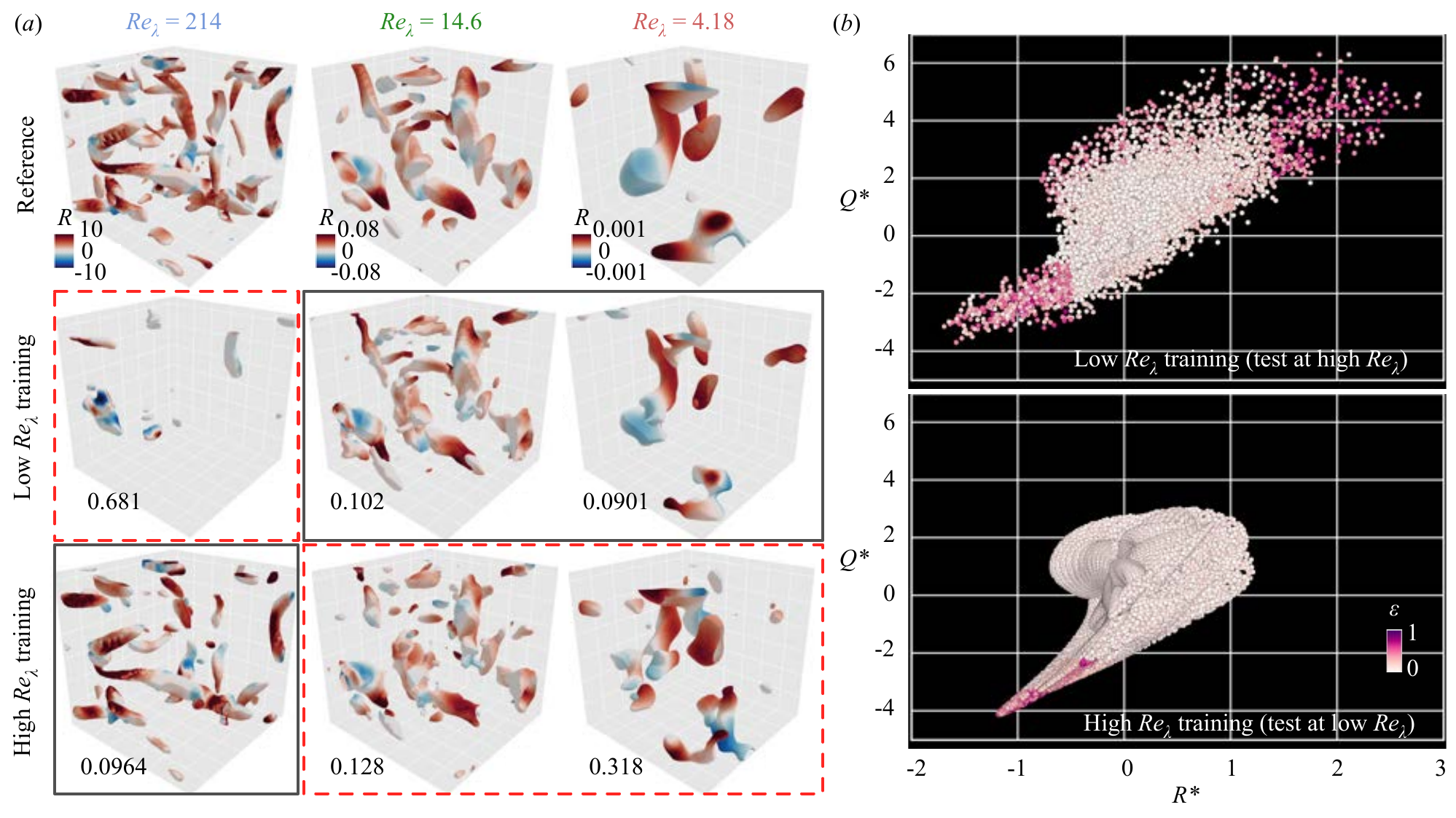}
    \vspace{-3mm}
    \caption{
    $(a)$ Super-resolution reconstruction of three-dimensional decaying turbulence.
    The reconstructed flow fields are visualized with the $Q$-criteria, colored by $R$.
    The values underneath each figure represent the $L_2$ error norm.
    The gray and red boxes respectively highlight snapshots for training and testing $Re_{\lambda}$ regime.
    $(b)$ Scaled $Q^*$ and $R^*$ at test $Re_{\lambda}$'s for low- and high-$Re_{\lambda}$ training cases, colored by the spatial $L_2$ reconstruction error.
    }
    \vspace{-6mm}    
    \label{fig5}
\end{figure}

The reconstructed flow fields are shown in figure~\ref{fig5}$(a)$.
The vortical flows are visualized using $Q$-criteria and colored by the third invariant $R$ computed from the reconstructed velocities.
For the low-$Re_{\lambda}$ training case, the model reconstructs the vortical structures for the training $Re_{\lambda}$ with an $L_2$ error of approximately 0.1.
In contrast, the visualized $Q$ variable at the high-$Re_{\lambda}$ exhibits significant level of error rendering the reconstruction grossly incorrect.
Such high errors can be explained using the scaled variables in figure~\ref{fig5}$(b)$ due to the flow features residing over the non-overlapping region on the present plane.  
The high error is also observed around the bottom left of figure~\ref{fig5}$(b)$.
This extrapolation region cannot be reconstructed with the low-$Re_{\lambda}$ training because these structures are barely seen in the low-$Re_{\lambda}$ data sets, as presented in figure~\ref{fig4}.

Next, let us consider the high-$Re_{\lambda}$ training case.
The reconstructed flow from the overlapping case (interpolation) is in agreement with the reference data.
In contrast to the extrapolatory low-$Re_{\lambda}$ training, reasonable reconstruction can still be achieved at $Re_{\lambda} = 4.18$.
This suggests that the high-$Re_{\lambda}$ training data holds insights into a wider range of vortical structures that also appear in the low $Re_{\lambda}$ regime, which is confirmed from the scaled $Q^*$ and $R^*$ data.
It is worth pointing out that the scaled invariants at the low $Re_{\lambda}$ include some non-overlapping portion (scale-variant structures) in the region of $Q^*<0$ and $R^*<0$.
This implies that the model trained with only the high-$Re_{\lambda}$ regime cannot cover vortical structures of $Q^*<0$ and $R^*<0$.
For better reconstruction over this regime, we need to include training data that covers $Q^*<0$ and $R^*<0$.
Such an observation provides guidance on how the vortical flow data should be prepared to enable reliable reconstruction.

\begin{figure}
    \centering
    \includegraphics[width=\textwidth]{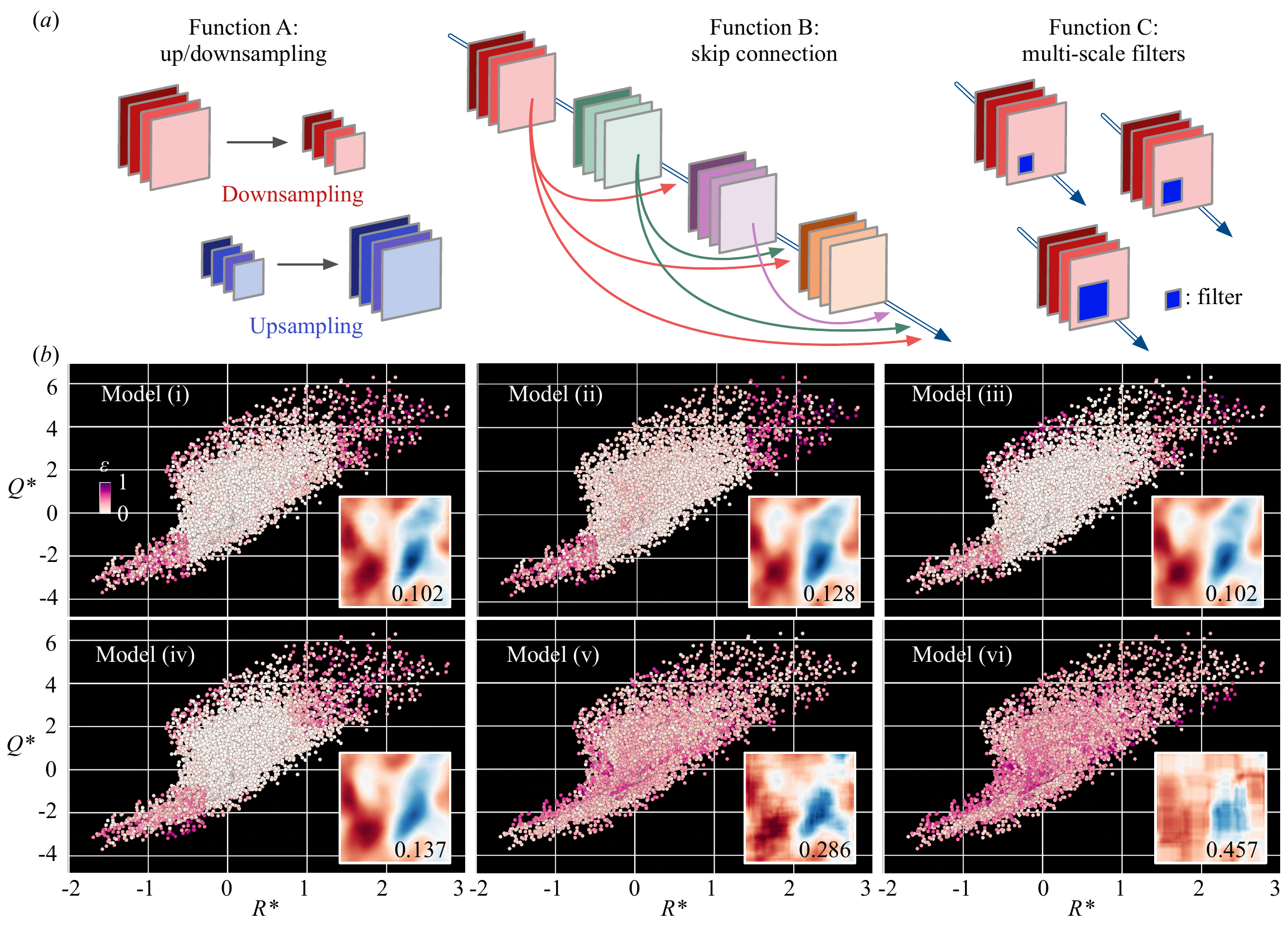}
    
    \caption{
    $(a)$ The three functions used in the DSC/MS super-resolution model.
    $(b)$ The scaled $Q^*$-$R^*$ data for {six} models.
    The two-dimensional sections of the reconstructed streamwise velocity are also shown with velocity errors.
    }
    \label{fig6}
\end{figure}

We further examine the nonlinearly scaled invariants to assess which inner functions of the machine-learning models contribute to robustness for scale-invariant regression.
We here consider {six} models using the aforementioned functions A, B, and C (figure~\ref{fig6}(a)): 
\begin{enumerate}
    \item the original DSC/MS model (functions A, B, and C), 
    \item {the original model without multi-scale filters (functions A and B)},
    \item the original model without the skip connection (functions A and C),
    \item the original model with up/downsampling only (function A),
    \item the original model with multi-scale filters only (function C),
    \item a regular convolutional neural network (without any of the functions above).
\end{enumerate}
These {six} models consider all possible combinations that can be constructed with the DSC/MS model.
We perform the low-$Re_{\lambda}$ training for all of these {six} cases.
The corresponding error distributions are presented in figure~\ref{fig6}$(b)$.
{For model (ii) which removes the multi-filter function from the baseline model, the error behavior on the scaled data is similar to that of the original model.
A similar observation is seen for models (iii) and (iv) which have no multi-filter functions.
Since models (ii-iv) include function~A, up/downsampling operations are crucial for robust super-resolution reconstruction of turbulent flows.}

This can be further confirmed with model (v) which is comprised only of the multi-scale filters (function C).
The reconstruction error is significantly higher without the up/downsampling operations.
Including dimension compression and expansion plays an important role in obtaining robustness for scale-invariant characteristics, which agrees with a number of studies on CNNs for scale invariance characteristics in image science \cite{jarrett2009best,van2017learning}.
In contrast, multi-scale filters have secondary importance in reconstructing the flow (model (v)) as evident from its comparison to the results from CNN (model (vi)).  
While the regular CNN without functions A, B, and C returns pixelized flow fields, model (v) achieves qualitative reconstruction.
The use of skip connections is also important for successful training to address issues related to the convergence of neural network weights~\cite{he2016deep}.
These findings suggest that robust turbulent flow reconstruction can be achieved by selecting machine-learning functions based on the implication of scaled invariants.

The current approaches to identify important model functions that accommodate scale invariance could be combined with emerging methods to investigate the role of networks in capturing flow physics, such as the sensitivity analysis \cite{lee2021analysis,jagodinski2023inverse} and Fourier analysis of the CNN filters \cite{subel2023explaining}.
Along with these techniques, model effectiveness and accuracy should be carefully examined in terms of its construction and the richness of training data for machine-learning applications of turbulent flows.

\section{Concluding remarks}
\vspace{-0.5mm}
\label{sec:conc}

\vspace{-0.5mm}
For nonlinear machine-learning models for turbulent flows, there is generally not a single parameter that can reveal whether such models are performing an interpolation or extrapolation. 
This is due to the turbulent flow data containing similarities in flow structures across a range of spatiotemporal scales.  
These properties contribute to machine-learning models being accurate beyond the coverage of training data in some cases.  
To shed light on the validity of machine-learning models, we nonlinearly scaled the invariants of the velocity gradient tensor $Q$ and $R$ with non-dimensional parameters using a Buckingham Pi theorem-based sparse regression, which maximizes the similarity of invariant data distributions over space and time across Reynolds numbers.  
As a canonical turbulent flow example, we considered three-dimensional decaying isotropic turbulence for $0.85 \leq Re_{\lambda} \leq 252$.  
The present approach found nonlinear scalings that express the influence of the decaying nature of the present flow, domain size of the simulation, and non-equilibrium effects of energy cascade on the invariants $Q$ and $R$.
With the scaled invariants $Q^*$ and $R^*$, we were able to determine that the training data lacked flow structures associated with strong biaxial strain and vortex stretching.

We further analyzed which types of machine-learning functions contribute to gaining robustness for scale-invariant vortical structures within the context of super-resolution reconstruction.
We found that fluid flow reconstruction can be achieved for data of the overlapping portion on the scaled $Q^*$--$R^*$ plane even under untrained $Re_{\lambda}$ with an appropriate construction of machine-learning models.
The present findings suggest that transfer learning could be effective for training nonlinear machine-learning models of turbulent flow across Reynolds number provided that extrapolatory structures do not alter the physics significantly \cite{inubushi2020transfer,guastoni2021convolutional}.
Including fractional exponents for Pi variables would likely enhance the generalizability of the present method for discovering additional nonlinear scaling in turbulent flows.
Incorporating nonlinearly scaled invariants into the training process could also support the generalizability of the models.
While we can consider the scaled invariants as the input and output of machine-learning models, they could also be incorporated into the loss function of machine-learning optimization.
The present procedure to examine nonlinear scalings of turbulent flow structures provides guidance on how to develop robust machine-learning models and compile the necessary training data, enabling us to depart from na\"ive training and being unaware of the validity of these complex models.

\section*{Acknowledgements}
{KF thanks the support from the UCLA-Amazon Science Hub for Humanity and Artificial Intelligence.
KT acknowledges the generous support from the US Air Force Office of Scientific Research (FA9550-21-1-0178) and the US Department of Defense Vannevar Bush Faculty Fellowship (N00014-22-1-2798).
DNS data was generously shared by Dr. M. Gopalakrishnan Meena.}

\section*{Declaration of interests}

{The authors report no conflict of interest.}


\bibliographystyle{unsrt}  
\bibliography{refs}

\end{document}